\documentclass[showpacs,preprintnumbers,twocolumn,prl,floatfix]{revtex4}
\usepackage{epsfig,graphicx,amsmath,amssymb,bm}

\begin{document}

\title{The sloppy model universality class and the Vandermonde matrix}
\date{\today}

\author{Joshua J. Waterfall$^1$, Fergal P. Casey$^2$, Ryan N. Gutenkunst$^1$,
  Kevin S. Brown$^3$, Christopher R. Myers$^4$, Piet W. Brouwer$^1$, Veit
  Elser$^1$, James P. Sethna$^1$}
\affiliation{$^1$ Laboratory of Atomic and
Solid State Physics, Cornell University,
Ithaca, NY 14853, USA}
\affiliation{$^2$ Center for Applied Mathematics, Cornell University, Ithaca, NY
  14853, USA}
\affiliation{$^3$ Department of Molecular and Cellular Biology, Harvard
  University, Cambridge, MA 02138, USA}
\affiliation{$^4$Cornell Theory Center, 
Cornell University, Ithaca, NY 14853, USA}

% XXX Check PACS
\pacs{02.30.Zz, 05.10.-a, 87.15.Aa, 87.16.Xa, 89.75.-k}

\begin{abstract}
In a variety of contexts, physicists study complex, nonlinear models with many
unknown or tunable parameters to explain experimental data.  We explain why such
systems so often are {\em sloppy}; the system behavior depends only on a few
`stiff' combinations of the parameters and is unchanged as other `sloppy'
parameter combinations vary by orders of magnitude. We contrast examples of
sloppy models (from systems biology, variational quantum Monte Carlo, and common
data fitting) with systems which are  not sloppy (multidimensional linear
regression, random matrix ensembles). We observe that the eigenvalue spectra for
the sensitivity of sloppy models have a striking, characteristic form, with a
density of logarithms of eigenvalues which is roughly constant over a large
range. We suggest that the common features of sloppy models indicate that they
may belong to a common universality class.  In particular, we motivate focusing
on a {\em Vandermonde ensemble} of multiparameter nonlinear models and show in
one limit that they exhibit the universal features of sloppy models.
\end{abstract}

\maketitle

\newcommand{\bigO}{{\mathcal{O}}}
\newcommand{\I}{{\mathrm{i}}}
\newcommand{\bp}{{\mathbf{p}}}
\newcommand{\ba}{{\mathbf{a}}}
\newcommand{\bg}{{\boldsymbol{\gamma}}}
\renewcommand{\d}{{\mathrm{d}}}
\newcommand{\CC}{\mathbb{C}}
\newcommand{\RR}{\mathbb{R}}
\newtheorem{theorem}{Theorem}
\newtheorem{corollary}{Corollary}
\newtheorem{conclusion}{Conclusion}
\newtheorem{conjecture}{Conjecture}
\newcommand{\lin}{{\mathrm{lin}}}

Systems with many parameters are often {\em sloppy}. For practical purposes
their behavior depends only on a few stiffly constrained combinations of the
parameters; other directions in parameter space can change by orders of
magnitude without significantly changing the behavior. Given a suitable cost
$C(\bp)$ measuring the change in system behavior as the parameters $\bp$ vary
from their original values $\bp^{(0)}$ (e.g., a sum of squared residuals), the
stiff and sloppy directions can be quantified as eigenvalues and eigenvectors of
the Hessian of the cost:
$
H_{ij} = \left.\partial^2 C/\partial p_i \partial p_j\right|_{\bp^{(0)}}.
$

\begin{figure}
\begin{center}
\includegraphics[width=8.4cm]{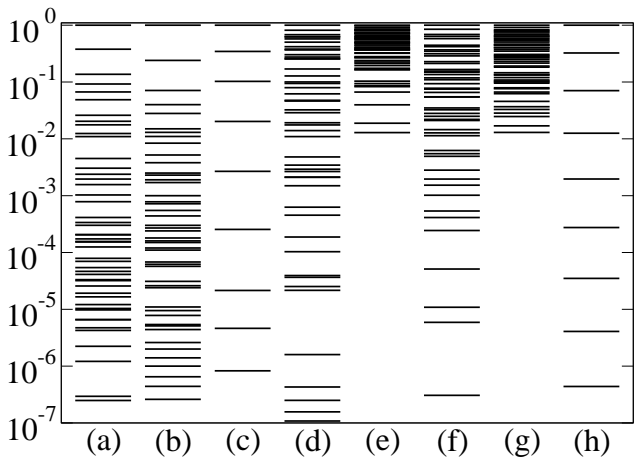}
\end{center}
\caption{
\label{fig:sloppyGallery}
Eigenvalues giving the stiffness/sloppiness of various models
as parameters are varied.  Each spectrum has been shifted so that the largest
eigenvalue is one.
%\hfill\break
(a) Growth factor signaling model (coupled nonlinear ODEs) for PC12
cells~\cite{kevinPC12}, as the 48 parameters (rate and Michaels-Menten
constants) are varied.
%\hfill\break
(b) Variational wave-function used in quantum Monte-Carlo, as the Jastrow
parameters (for electron-electron coincidence cusps) are varied,
%\hfill\break
(c) Radioactivity time evolution for a mixture of twelve common radionuclides as
 the half-lives $\gamma_i$ are varied.  The radionuclides are those available
 from Perkin-Elmer~\cite{PerkinElmer} with half-lives less than 100 days. (Only
 the first nine eigenvalues are shown.)
%\hfill\break
(d) The same exponential decay model as in (c) with 48 decay constants
 $\gamma_i$ randomly spread over a range of $e^{50}$.
%\hfill\break
(e) One random 48$\times$48 matrix in the Gaussian Orthogonal Ensemble (GOE)
(not sloppy).
%\hfill\break
(f) A product of five random 48$\times$48 matrices, illustrating the random
 product ensemble (not sloppy, but ill conditioned).
%\hfill\break
(g) A plane in 48 dimensions fit to 68 data points, the same number and data
points as for the biology model in column~(a) (Wishart statistics, not sloppy). 
%\hfill\break
(h) A polynomial fit to data, as the 48 monomial coefficients are varied (the
Hilbert matrix~\protect\cite{Hilbert}, sloppy).}
\end{figure}

Figure~\ref{fig:sloppyGallery} shows the eigenvalues of the cost Hessian for
many different systems; those in (a), (b), (c), (d) and (h) are all sloppy.
The sensitivity of model behavior to changes along an eigenvector is given by
the square root of the eigenvalue---therefore the range in eigenvalues of
roughly one million for the sloppy models means that one must change parameters
along the sloppiest eigendirection a thousand times more than along the
stiffest eigendirection in order to change the behavior by the same amount.
Although anharmonic effects rapidly become important along sloppy
eigendirections, a principal component analysis of a Monte-Carlo sampling of
low-cost states has a similar spectrum of eigenvalues \cite{kevinSloppyPRE}; the
sloppy eigendirections become curved sloppy manifolds in parameter space.
Similar sloppy behavior has been demonstrated in fourteen systems biology models
taken from the literature~\cite{kevinPC12, Ryan}, and in three multiparameter
interatomic potentials fit to electronic structure data~\cite{SloppyMo}.  In
these disparate models we see a common, peculiar behavior: the $n$th stiffest
eigendirection is more important than the ($n{+}1$)th by a roughly constant
factor, giving a total range of eigenvalues of typically over a million for any
model with more than eight parameters. We call systems
exhibiting these characteristic features {\em sloppy models}.

This sloppiness has a number of important implications.  In estimating
prediction errors, sloppiness affects both the estimation of statistical errors
due to uncertainties in the experimental data~\cite{kevinPC12,Ryan} and allows
an estimation of systematic errors due to imperfections in the models (for
example in interatomic potentials~\cite{SloppyMo} and density functional
theory~\cite{SloppyDFT}).  It makes extracting parameter values from fits to
sloppy models ill-posed~\cite{kevinPC12,Varner}.  Conversely, it is much more
efficient to improve the predictivity of a model by fitting parameters to system
behavior than by designing experiments that precisely determine the individual
parameter values~\cite{Ryan}.  Sloppy problems are also better approached with
optimization algorithms~\cite{NumericalRecipes,FergalSMC} (like the
Levenberg--Marquardt and Nelder--Mead methods) which can adapt to widely
diverging step sizes along different parameter combinations.

Let us begin with the famously ill posed problem of fitting a sum of
exponentials to data \cite{lanczos, vdbos93}. Consider a mixture of equal
amounts of $N$ radioactive elements whose decay signal is thus the sum of $N$
exponentials with decay rates $(\gamma_1^{(0)},
\ldots,\gamma^{(0)}_N)$.  We define a cost function for general decay rates as
$C = \frac{1}{2} \int_0^\infty ( \sum_{i=1}^N \exp(-\gamma_i t) -
\sum_{i=1}^N \exp(-\gamma_i^{(0)} t))^2 \d \log t$ (spacing the `data points'
equally in log time makes analyzing large ranges of decay constants convenient).
Because the decay constants are positive and can have a large range of sizes, we
use their logarithms as our parameters ($p_i = \log \gamma_i$), giving model
sensitivity to relative changes in the decay rates. The resulting Hessian is
$H_{ij}|_{\bp^{(0)}} = 2 \gamma^{(0)}_i \gamma^{(0)}_j/(\gamma^{(0)}_i +
\gamma^{(0)}_j )^2$.  For the twelve radionuclides described in the caption to
Figure~\ref{fig:sloppyGallery} (c), the eigenvalues of the Hessian are each
separated by nearly one decade; the sloppiest mode has an eigenvalue a factor of
$10^{10}$ smaller (less important) than the stiffest.  This is not the result
of an inaccurate mathematical description, it is true for the correct
model and parameters with a complete complement of data.  The origin of
sloppiness is not a simple lack of data where trivial overparameterization leads
to unidentifiable parameters.

Unless the individual lifetimes are well separated, the net radiation cannot be
used to measure the lifetimes reliably.  The difficulty is that the signal is
the sum of many functions with similar shapes; one can generate almost identical
signals with wildly different values for the parameters.  Similarly, the
sloppiness in more realistic models is presumably due to the compensation of
subsets of parameters with similar effects.
If we pick 48 lifetimes whose
logarithms are instead uniformly distributed over a range of $2\epsilon = 50$
(largest/smallest $\approx \mathrm{e}^{50} \approx 10^{21}$), the density of
levels and the variation
in spacings between neighboring levels in the new spectrum
(Figure~\ref{fig:sloppyGallery}(d)) is similar to that of the real-life
models in \ref{fig:sloppyGallery}(a) and~(b).

While a large number of models are sloppy, not all multiparameter models share
this quality.  The simplest form of multiple linear regression, which is in
essence fitting a plane through the origin to a cloud of points, is not sloppy.
The Hessian matrix for this type of model is the sample covariance matrix of the
data points and is known as a Wishart matrix \cite{Wishart}.  The eigenvalues of
a Wishart matrix are described by the
Mar\v{c}enko-Pastur distribution 
\cite{MarcenkoPastur} and an example is seen in
Figure~\ref{fig:sloppyGallery}(g).  The classic ensembles of Random Matrix
Theory \cite{mehtabk, RMTreviewNEW} (Figure~\ref{fig:sloppyGallery}(e)) have
uniform eigenvalue densities instead of the exponentially large range
characteristic of sloppy systems.  The ensemble of products of random
matrices~\cite{beenakkerRMT} (Figure~\ref{fig:sloppyGallery}(f)) does mimic the
exponential spacing of (singular) values but in this case the variance of level
spacings is proportional to the mean spacing.  Toward the end of this paper we
will see that for sloppy models, in the limit of large spacings, the variance is
instead independent of the mean.

Why are so many models sloppy? We can gain insight by considering fitting data
for $x\in [0,1]$ with polynomials. If one considers the polynomials of
order $N$ to be sums of monomials, $y(x,\bp) = \sum_{i=0}^N p_i x^i$, 
the Hessian
is $H_{ij} = 2 A_N = \frac{2}{i+j+1}$, the famously ill-conditioned Hilbert
matrix (Figure~\ref{fig:sloppyGallery}(h)).
Indeed, the coefficients of the monomials are known to be poorly determined in
such polynomial fits~\cite{NumericalRecipes}.
Suppose we instead generate the same polynomial fit, but parameterize our
polynomial as a sum of the appropriate shifted Legendre polynomials 
$y(x,\bp') = \sum_{i=0}^N p'_i L_i(x)$; $L_0 = 1$, $L_1 = \sqrt{3}(2x - 1)$,
$L_2 = \sqrt{5}(6x^2 - 6x + 1)$, \dots.
The shifted Legendre polynomials
are orthonormal in the $L^2$ norm on $[0,1]$, and the Hessian in the $\bp'$
basis is the identity matrix. By changing our parameterization from monomial
coefficients $\bp$ to coefficients $\bp'$ in the appropriate orthonormal basis,
our sloppiness is completely cured. The sloppiness is due to the fact that
the monomial coefficients (natural from many perspectives) are a perverse set of
coordinates from the point of view of the behavior of the resulting
polynomial. We can quantify this by noting that the transformation $S_N$ from
the monomial basis to the orthonormal basis (the coefficients of the shifted
Legendre polynomials) has a tiny determinant, and therefore the
volume enclosed by the monomial basis vectors shrivels and becomes greatly
distorted under the transformation.  This determinant can be found by
noting that $S_N$
gives a Cholesky decomposition of the Hilbert matrix $A_N = S_N^\top S_N$ 
, and thus $\det S_N = \sqrt{\det A_N} =
(\prod_{i=1}^{N-1}(i!))^2/\sqrt{\prod_{j=1}^{2N-1}(j!)}$ ~\cite{Hilbert}.
Physically, the monomials all
have roughly the same shape (starting flat near zero, and rising sharply at
the end near one), and can be exchanged for one another, while the orthogonal
polynomials all have quite distinct shapes. In nonlinear sloppy models
the sloppiness is more difficult to remove: (a)~the 
transformation to unsloppy parameters will be nonlinear away from the optimum,
often not even single-valued, (b)~we may
not have the insight or the ability to change parameterizations to those
natural for fitting purposes, and (c)~often the natural parameterization
is determined by the science (as in biochemical rate constants, arbitrary
linear combinations of which are not biologically motivated).

What causes this even distribution of relative
stiffnesses over so many decades of scales?
To form strong conclusions about sloppy models we must establish criteria
sufficient to exclude the large variety of multiparameter systems that
will not be sloppy.
First, we specialize to models where the cost is a sum of squared residuals 
$C(\bp)=\sum_m r_m^2$, where the sum may be continuous 
(e.g., an integral over time) and $r_m = y_m(\bp)-d_m$ is the deviation of
theory $y(\bp)$ from the experimental datum $d_m$; all of our examples
of sloppy models are of this type. Second, to avoid including systems where
each parameter is the subject of a separate experiment isolating that component,
we make the (strong) assumption that all of the residuals $r_m(\bp)$ depend on
the parameters $\bp$ in a symmetric fashion (e.g., permuting $p$ leaves $r_m$
unchanged).  This allows us to recast the residuals into the basis of power sum
polynomials of the parameters, $r_m(\mu_1, \mu_2, \dots), \mu_k = \sum_{i=1}^N
p_i^k$, which can also be viewed as the moments of the parameter distribution.  
%Permutation
%symmetry is obeyed by our fitting exponentials problem
%but is violated by polynomial fits and the real world systems. We have seen,
%however, that the different polynomials have almost equivalent shapes 
%in fitting the data and that this similarity is likely the source of
%sloppiness. In the biological and variational wave function examples,
%many of the basis functions are also quite similar in functional form. 
Third, we noticed that in fitting exponentials the compensable nature of
different parameters increased when they were restricted to smaller ranges; 
here we will assume that the parameters are all confined to a small range 
$p_i \in [\bar{p} \pm \epsilon]$. Thus if we define $\epsilon_i = p_i -
\bar{p}$, the residuals $r_m(\mu_1, \mu_2, \dots)$ can be written as functions
of the moments $\mu_k = \sum_{i=1}^N \epsilon_i^k$.

%\begin{conclusion}\label{VTATAV}
%For a cost function which is a sum of squared residuals
%  $C(\bp)=\sum_m r_m^2$, if each residual $r_m(\bp)$ is a symmetric function
%  of the parameters ${p_1,...,p_N}$ and if the parameters are confined to a 
%  range $p_i^{(0)} \in [\bar{p} \pm \epsilon]$, then the Hessian matrix 
%  $H_{ij}=\left.\partial^2 C/\partial p_i \partial p_j\right|_{\bp^{(0)}}$
%  can be decomposed into
%\begin{equation}
%H = V^\top A^\top A V
%\end{equation}
%where the elements of $A$ are bounded as $\epsilon \to 0$ and $V$ is the
%Vandermonde matrix, $V_{kj} = \epsilon_j^{k-1}$.

%\noindent  
%{\em 

In general the Hessian is
\begin{equation}
H_{ij} = \sum_{m} \left(\frac{\partial r_m}{\partial p_i} \frac{\partial
  r_m}{\partial p_j} + r_m \frac{\partial^2 r_m}{\partial p_i \partial
  p_j}\right)
\end{equation}
but for the correct model at the true parameters the cost is zero, so $r_m = 0
\; \forall m$ and $H = J^\top J$ with the Jacobian
\begin{equation}
J_{mj} = \frac{\partial r_m}{\partial p_j} 
= \sum_{k=1}^K \frac{\partial r_m}{\partial \mu_k} k \epsilon_j^{k-1}
= A_{mk}V_{kj}\\
\end{equation}
%\begin{equation}
%A_{ik} = \frac{\partial r_i}{\partial m_k} k\\
%\end{equation}
%\begin{equation}
%V_{kj} = \epsilon_j^{k-1}
%\end{equation}
where $A_{mk} = \frac{\partial r_m}{\partial \mu_k} k$, $V_{kj} =
\epsilon_j^{k-1}$, and $K$ is the maximum degree (possibly $\infty$) to which 
we expand in $\epsilon$.  Thus $H = J^\top J = V^\top A^\top A V$. 
%\hfill $\Box$}
%\end{conclusion}
%
Here $V$, the famous Vandermonde matrix, 
is the heart of the sloppy model universality class.  
%The Vandermonde matrix does appear routinely in least squares polynomial
%fitting
%to separate the linear parameters from the nonlinear dependent variables but
%its 
%role here in the Hessian of nonlinear parameterizations is distinct.
Reminiscent of random
matrix theory ensembles, we are now interested in the Vandermonde ensemble of
Hessians of the form $V^\top A^\top A V$.
The Vandermonde matrix is
well-known primarily because its determinant (for $N=K$) can be expressed
analytically,
$\det(V) = \prod_{i<j} (\epsilon_i - \epsilon_j)$.  As $\epsilon \to 0$ this
product is tiny,  $\det(V) = \bigO(\epsilon^{N(N-1)/2})$. 
While the elements of $A$ do, in general, depend on the parameter values, they
either approach a constant or zero in this limit and we can see that the
determinant of $H$, $\det(H) = \det(V)^2 \det(A)^2$ is smaller still.  As we saw
with the Hilbert matrix and fitting monomials to data, transformation matrices
with very small determinants are a signature of sloppy models.

To show that the eigenvalues in our Vandermonde ensemble are evenly spread in
logarithm, we will make use of an apparent truth about matrices:
\begin{conjecture}\label{GeneralEigBounds}
Let $S \in \RR^{n\times n}$ be symmetric and positive definite.  Let $E \in
\RR^{n\times n}$ be diagonal with $E_{ii} = \epsilon^{i-1}$ and
$0<\epsilon\ll 1$.  Then the $m$th largest eigenvalue of $ESE$ is
$\bigO(\epsilon^{2(m-1)})$ (less than some constant times $\epsilon^{2(m-1)}$).

\noindent
{\em We have two reasons to believe this conjecture is true.
(1) Treating the off-diagonal components of $ESE$ as a perturbation, the
corrections to the $m$th eigenvalue are of order $\epsilon^{2(m-1)}$ to all
orders in perturbation theory, despite the fact that many of the perturbing
elements are large compared to the diagonal entries.
(2) Extensive numerical tests show an even sharper result: the $m$th largest
eigenvalue, $\lambda_m$, is bounded above by the $m$th largest row sum of $EES$
for all
$\epsilon$, where the row sum for row $k$ is $\sum_l \epsilon^{2(k-1)} \vert
S_{kl} \vert$.  This implies that $\lambda_m \leq \Vert S \Vert_\infty
\epsilon^{2(m-1)}$, and also (for $\epsilon = 1$) implies the remarkable
apparent fact that the sorted eigenvalues of any symmetric positive definite
matrix are each bounded by their corresponding sorted row sums.
\hfill $\Box$}
\end{conjecture}

%We have numerical evidence that to leading order in $\epsilon$ the
%eigenvectors of the Hessian are the right singular vectors of the Vandermonde
%matrix. (A non-square matrix $V$ has a singular value decomposition 
%$V=U \Sigma W^\top$, where the non-square matrix $\Sigma$ has the singular
%values
%of $V$ along the diagonal and is zero elsewhere; $W$ gives the right singular
%vectors, which are eigenvectors of $V^\top V$, and $U$ gives the left singular
%vectors, eigenvectors of $V V^\top$.)
%Motivated by this, we shall transform our Hessian to this right singular basis
%of 
%$V$. 

Motivated by numerical evidence that to leading order in $\epsilon$ the
eigenvectors of the Hessian are the right singular vectors of the Vandermonde
matrix, we shall transform into that basis.
We first bound the singular values of the Vandermonde matrix. Conveniently,
$V V^\top$ has the form necessary for Conjecture \ref{GeneralEigBounds}.
%
%\begin{conclusion}\label{VandermondeSVD}
%The $m$th-largest singular value 
%$\sigma_m(V)$ of the $N$-column Vandermonde matrix,
%  $V_{ij} = \epsilon_j^{i-1}$ 
%is $\bigO(\epsilon^{m-1})$.
%
%\noindent  
%{\em 
The singular values of $V$ are the positive square root
of the eigenvalues of $V V^\top$.  Factoring the appropriate power of $\epsilon$
from each row of the Vandermonde matrix gives $V = E X$ and $V V^\top = E X
X^\top E$ where $E$ is the same as in Conjecture \ref{GeneralEigBounds} and the
elements of $X$ are bounded by one. Equating $X X^\top$ with the matrix $S$ in
Conjecture \ref{GeneralEigBounds}, we conclude that the eigenvalues of $V
V^\top$ scale as $\lambda_m(V V^\top) = 
\bigO(\epsilon^{2(m-1)})$ and thus $\sigma_m(V) = \bigO(\epsilon^{m-1})$.
%\hfill $\Box$}
%\end{conclusion}

We now transform the Hessian into this basis, and again use Conjecture
\ref{GeneralEigBounds} to bound its eigenvalues.
%
%\begin{conclusion}\label{Main} 
%The eigenvalues of the Hessian matrix for the class of models in Conclusion
%\ref{VTATAV} scale as
%\begin{equation}
%\lambda_i(H) = \bigO(\epsilon^{2(i-1)})
%\end{equation}
%
%\noindent
%{\em 
%We have already derived $H = V^\top A^\top A V$.  
Starting with the decomposition $H = V^\top A^\top A V$, taking the
  singular value decomposition of $V = U \Sigma W^\top$, and transforming the
  Hessian into the basis of the right singular vectors of the $V$, we have
  $W^\top H W = \tilde{H} = \Sigma^\top U^\top A^\top A U \Sigma$.  %By
  %Conclusion  \ref{VandermondeSVD}
We know that $\Sigma_{ii} = \bigO(\epsilon^{i-1})$.  By
  construction the elements of $A$ are well-behaved as $\epsilon \to 0$
  and since $U$ is an orthogonal matrix its elements too cannot diverge in this
  limit.  This means that $\tilde{H}_{ij} = \bigO(\epsilon^{i+j-2})$.  By
  Conjecture  \ref{GeneralEigBounds} we know that $\lambda_i(\tilde{H}) =
  \bigO(\epsilon^{2(i-1)})$ and since $\tilde{H}$ is simply an orthogonal
  transformation of $H$, $\lambda_i(H) = \bigO(\epsilon^{2(i-1)})$.
%\hfill $\Box$}
%\end{conclusion}
%While rigorous universality is only expected as the system size approaches
%infinity, we find empirically that models with more than roughly eight
%parameters are often recognizably sloppy~\cite{Ryan}.
Rigorous universality is only expected as the system size approaches infinity.
Empirically we find, from studying a variety of models~\cite{Ryan, SloppyMo,
  SloppyDFT} as well as subsystems of models (like PC12 in
Figure~\ref{fig:sloppyGallery} (a))~\cite{JoshThesis}, that models with 
more than roughly eight parameters are often recognizably sloppy.

Do these results tell us anything about the statistics of level spacings?
Unless two parameters are strictly
degenerate or the residuals are independent of a particular moment of the
parameter distribution, $\lambda_i = l_i \epsilon^{2(i-1)}$ for some non-zero
coefficient $l_i$.  The relative spacing between neighboring eigenvalues is $s_i
= \log(\lambda_i/\lambda_{i+1}) = \log(l_i/l_{i+1}) - 2\log \epsilon$.  For a
fixed model but an ensemble of random parameters, the distribution of
coefficients $l_i$ has a finite width as $\epsilon \to 0$.  Therefore the
distribution of $s_i$ over the ensemble, normalized by $2\log \epsilon$
such that the average spacing is unity, goes to one with a width which vanishes
as $\epsilon \to 0$.  This means that the whole system is becoming not only more
sloppy (larger spacing) but it is becoming almost deterministically so (strong
level repulsion).  Figure~\ref{fig:sloppyGallery} (c) is a clear depiction of
this remarkably strong level repulsion.

%What is the link between the Vandermonde ensemble at small $\epsilon$ and the
%behavior of real world sloppy models (Figure~\ref{fig:sloppyGallery} columns
%(a), (b)) and also the behavior at large $\epsilon$ (column
%(d))?  These systems share the roughly uniform density of log-eigenvalues over
%many decades that is the signature of sloppy models, but do not exhibit strong
%level repulsion.  We picture these systems
%as weakly-coupled collections of our small-$\epsilon$ Vandermonde ensembles.
%Thus we view the fastest and slowest decay rates for the model in column (c) as
%nearly decoupled, with internal strong level repulsion superimposed to
%produce broad but Poisson level statistics.  Similarly, we view the
%parameter space for a real world model (column (a) or (b)) as decoupling
%into weakly-coupled subspaces, inside of which all parameter directions lead to
%nearly equivalent changes in model behavior.

What is the link between the Vandermonde ensemble at small $\epsilon$ and the
behavior of real world sloppy models (Figure~\ref{fig:sloppyGallery} columns
(a), (b)) and also the behavior at large $\epsilon$ (column (d))?  These systems
share the roughly uniform density of log-eigenvalues over many decades that is
the signature of sloppy models but do not exhibit strong level repulsion.  The
real world models also do not share the strict requirement that the residuals be
perfectly symmetric functions of the parameters.  We conjecture that while
not all of the parameters are interchangeable in real world sloppy models, there
are Vandermonde subsystems lurking below the surface.  Thus the fastest decay
rates in column (d) constitute one Vandermonde subsystem and the slowest decay
rates another.  Indeed, the Poisson statistics of level spacings when fitting
exponential decays from a wide range (e.g. $2\epsilon = 50$ as in (d)) can be
reproduced by superimposing the spectra of several separate experiments, each
fitting decays from a narrower range (e.g. $2\epsilon = 3.5$ as in (c)).  Such a
decomposition into Vandermonde subsystems is also illustrated by modifying the
net radiation model to include the initial amounts of the elements as unknown
parameters.  Now the parameters clearly separate into two classes -- decay rates
and initial amounts.  Each class alone fits the assumptions of the Vandermonde
ensemble, produces rigidly (strong level repulsion) sloppy spectra, and
generates nearly equivalent patterns of changes in the residuals.  When mixed
together however, the fact that parameters from one class can not compensate for
parameters of the other class destroys the correlations between levels and they
do not repel each other anymore.  Similarly, a full many body wave function in
quantum Monte Carlo~\cite{CyrusBook} decomposes into the sloppy space of the
Jastrow parameters in figure~\ref{fig:sloppyGallery} (b) and a non-sloppy
subspace of the Configuration Interaction coefficients describing
single-particle orbitals.

These results motivate algorithms for the
decomposition of real world sloppy models into rigidly sloppy Vandermonde
subspaces whose components are effectively redundant.  Such a
decomposition would be useful for three separate reasons: a) explaining why a
particular model is sloppy overall, b) suggesting routes for model
reduction and coarse graining by subsuming degrees of freedom within Vandermonde
systems, and c) prescribing changes in parameters to alter specific aspects
of model behavior.

Complex models from a wide array of scientific fields are {\em sloppy}: they
each have an exponentially large
range of sensitivities to changes in underlying parameter values.  This occurs
because the parameters natural
for experimental manipulation or human description are often a severe
distortion of the basis natural for describing system behavior.  Far from being
a deficiency, sloppiness is
in fact a saving grace of complex models---provided the right combinations of
parameters are known they provide nontrivial and
well-constrained predictions despite surprisingly unconstrained parameters
overall.  Understanding the origins and implications of sloppiness in its
various incarnations offers new, fundamental insights into complex systems.

\begin{acknowledgments}
The authors thank R.~Cerione, P. Flaherty, D. Schneider, and C. Umrigar for
helpful discussions.  We also thank C. Umrigar for the variational quantum Monte
Carlo eigenvalues.  JJW was supported by a DOE Computational Science Graduate
Fellowship.  RNG was supported by an NIH Molecular Biophysics Training Grant,
T32-GM-08267.  CRM acknowledges support from USDA-ARS project 1907-21000-017-05.
This work was supported by NSF grant DMR-0218475.
\end{acknowledgments}

\end{document}